# Mechanical interpretation of the Klein-Gordon equation

Valery P. Dmitriyev

*Lomonosov University*
*P.O.Box 160 Moscow 117574 Russia*
*e-mail: dmitr@cc.nifhi.ac.ru*

The substratum for physics can be seen microscopically as an ideal fluid pierced in all directions by the straight vortex filaments. Small disturbances of an isolated filament are considered. The Klein-Gordon equation without mass corresponds to elastic stretching of the filament. The wave function has the meaning of the curve's position vector. The mass part of the Klein-Gordon equation describes the rotation of the helical curve about the screw axis due to the hydrodynamic self-induction of the bent vortex filament.

## Vortex filament

We consider the motion of an isolated vortex filament in the ideal fluid. The vortex filament goes as a microscopic constituent of the substratum for physics historically referred to as the vortex sponge. The latter is usually seen as the ideal fluid pierced randomly in all directions by the straight vortex filaments. We are interested in small perturbations of the filament. There are two kinds of them. In stretching deformation the filament behaves as an elastic string. Because of the hydrodynamic self-induction the bent vortex filament evolves changing the form and position in the space.

Let the filament be directed along the $x$ axis. Then the small perturbation of the filament can be specified considering the position vector $\mathbf{r}$ as a function of $x$

$$\mathbf{r}(x,t) = x\mathbf{i}_1 + y(x,t)\mathbf{i}_2 + z(x,t)\mathbf{i}_3 \tag{1}$$

where the dependence on the time $t$ is also included.

## Elasticity

In small stretching deformation a vortex filament behaves as an elastic string. The motion of an elastic string is governed by the d'Alembert equations

$$\begin{aligned}\frac{\partial^2 y}{\partial t^2} &= c^2 \frac{\partial^2 y}{\partial x^2} \\ \frac{\partial^2 z}{\partial t^2} &= c^2 \frac{\partial^2 z}{\partial x^2}\end{aligned} \tag{2}$$

Commonly that describes a plane wave propagating along the string with the speed $c$:

$$af(kx - \omega t), \quad \omega = ck$$

Taking the phase shift for two sinusoidal transverse waves as $\pi/2$, a circularly polarized wave can be constructed:

$$\begin{aligned}y &= a\cos(kx - \omega t) \\ z &= a\sin(kx - \omega t)\end{aligned}$$

This wave has the shape of a helix with the wave number $k$ taking the meaning of the curve's torsion $\tau$. The longitudinal motion of the helix looks as the rotation about the $x$ axis with the angular velocity $\omega$. The circular wave can be conveniently expressed in the complex valued form

$$\varphi(x,t) = a\exp[i(kx - \omega t)] \tag{3}$$

It obeys the Klein-Gordon equation without mass

$$\frac{\partial^2 \varphi}{\partial t^2} = c^2 \frac{\partial^2 \varphi}{\partial x^2} \qquad (4)$$

## Self-induction

The fluid element of the bent vortex filament moves in the space due to the hydrodynamic interaction of the adjacent elements. In the local induction approximation the motion of the filament is described by the equation

$$\frac{\partial \mathbf{r}}{\partial t} = \nu \frac{\partial \mathbf{r}}{\partial l} \times \frac{\partial^2 \mathbf{r}}{\partial l^2} \qquad (5)$$

where $\nu = \text{const}$ and $l$ the length measured along the filament. For small perturbations: $l \approx x$. Using it and (1) in (5) gives [1] neglecting quadratic terms

$$\frac{\partial \mathbf{r}}{\partial t} = \nu \, \mathbf{i}_1 \times \left( \frac{\partial^2 y}{\partial x^2} \mathbf{i}_2 + \frac{\partial^2 z}{\partial x^2} \mathbf{i}_3 \right)$$

The latter can be conveniently rewritten in the complex valued form

$$\frac{\partial \varphi}{\partial t} = i\nu \frac{\partial^2 \varphi}{\partial x^2} \qquad (6)$$

where

$$\varphi = y(x,t) + iz(x,t)$$

The Schroedinger equation (6) is met by the helical curve

$$\varphi(x,t) = a \exp\left[ i \left( \tau x - \nu \tau^2 t \right) \right] \qquad (7)$$

where $\tau = \text{const}$ is the torsion of the helix and the curvature is given by $\kappa = a\tau^2$. The helix rotates around the $x$ axis with the angular velocity $\nu \tau^2$.

## Klein-Gordon equation

The motion of a stretched vortex filament combines both the self-induction and the elasticity. For an isolated filament the solution must have the form of a helix (7), or (3), though a correction to the frequency should be made.

Notice that the left-hand part of the dynamic equation (4) has the meaning of the acceleration and the right-hand part – of the force. Differentiating (7) twice with respect to the time $t$ we find the acceleration in the self-induction circular motion

$$-\nu^2 \tau^4 \varphi$$

It is just the quantity, which the elastic equation (4) should be improved with:

$$\frac{\partial^2 \varphi}{\partial t^2} = c^2 \frac{\partial^2 \varphi}{\partial x^2} - \nu^2 \tau^4 \varphi \qquad (8)$$

Next, we will express the coefficient $\nu^2 \tau^4$ from (8) in physical terms. In this event the relations found earlier in [1] will be taken into account. First, the soliton on a vortex filament moves along the $x$ axis with the velocity [2]

$$\upsilon = 2\nu\tau$$

Looking otherwise, the latter is the group velocity for the small amplitude self-induction wave on a vortex filament [1]. In order to use it in (8) we choose for $\upsilon$ the maximal value $c$. As was shown in [1], the less the amplitude $a$ of the soliton the greater is its translational velocity $\upsilon$. That implies the concept of the minimal amplitude given by

$$a_0 = 2\nu / c$$

and the corresponding elementary helix. The notions of the soliton's mass, momentum and the energy were also introduced. It was found for the mass

$$m_0 = \varsigma a_0$$

where $\varsigma$ is the linear density of the fluid along the filament. Comparing the mechanical model of the wave-particle with the standard description there was found [1] for the Planck constant:

$$\hbar = 2\nu\varsigma a_0$$

Now, using the above relations in (8) we come to the standard form of the Klein-Gordon equation

$$\frac{\partial^2 \varphi}{\partial t^2} = c^2 \frac{\partial^2 \varphi}{\partial x^2} - \frac{m_0^2 c^4}{\hbar^2} \varphi \qquad (9)$$

Insofar as a quantum particle can be seen [1] as a dispersion of, say, $m$ elementary helices, the linear equation (9) is easily extended to a larger mass $m_\varepsilon$ using the relation

$$m_\varepsilon = m\varsigma a_0$$

Take notice that the equation (2) is valid only for very small stretching of the filament. So, in order to model a mass particle some extra segment $ma_0$ of the vortex filament needs to be added.

## Maxwell's equations

The equation for the electromagnetic wave is a trivial consequence of (2). In the elastic model [3] the magnetic vector potential corresponds to the displacement field of the quasisolid aether i.e.

$$A_1 \propto y \;, \; A_2 \propto z$$

In the turbulent aether it is modeled [4] by the perturbation of the average fluid velocity:

$$\mathbf{A} \propto \delta\langle\mathbf{u}\rangle$$

or rather by the very average velocity since the background value is vanishing in the averaging. The perturbation of the fluid velocity field due to the torsional wave on the vortex filament can be found substituting (1) with (3) in the Biot-Savart law for the hydrodynamic induction:

$$\delta\mathbf{u} = \frac{\Gamma}{4\pi}\int_l \frac{d\mathbf{r}\times\mathbf{s}}{s^3} - \frac{\Gamma}{4\pi}\int_x \frac{d\mathbf{x}\times\mathbf{s}}{s^3} \approx \frac{\Gamma}{4\pi}\int_x \frac{d\vec{\varphi}\times\mathbf{s}}{s^3} \qquad (10)$$

where $\Gamma$ is the circulation of fluid velocity around the filament, $\mathbf{s}$ the radius vector from a point on the filament to the point in the fluid, $h$ the least distance from the latter to the filament. Here the complex wave function (3) was treated as a vector quantity. For $ka \ll 1$ (10) can be evaluated as

$$\delta\mathbf{u} \propto \frac{\Gamma}{4\pi}\frac{\partial\varphi}{\partial x}\int_x \frac{d\mathbf{x}\times\mathbf{s}}{s^3} = i\frac{\Gamma k}{4\pi h}\varphi$$

where the velocity was treated as a complex value.